\documentclass[prl,amsmath,amssymb,dvips]{revtex4}
\usepackage{graphics,epsfig,amssymb}

\def\beq{\begin{equation}} 
\def\eeq{\end{equation}} 
\begin{document}

\title{Quartet correlations in N=Z nuclei induced by  realistic two-body  interactions}

\author{M. Sambataro$^a$\footnote{e-mail: michelangelo.sambataro@ct.infn.it}
and  N. Sandulescu$^b$\footnote{e-mail: sandulescu@theory.nipne.ro}} 
\affiliation{$^a$Istituto Nazionale di Fisica Nucleare - Sezione di Catania,
Via S. Sofia 64, I-95123 Catania, Italy \\
$^b$National Institute of Physics and Nuclear Engineering,  76900 Bucharest-Magurele, Romania}

\begin{abstract}
Two variational quartet models previously employed in a treatment of pairing forces  are extended to the case of a general two-body interaction. One model approximates the nuclear states as a condensate of identical quartets  with angular momentum $J=0$ and isospin $T=0$ while the other let these quartets to be all different from each other. With these models we investigate the role of alpha-like quartet correlations both in the ground state and in the lowest $J=0$, $T=0$ excited states  of  even-even $N=Z$ nuclei in the $sd$-shell.
We show that the ground state correlations  of these nuclei can be described
to a good extent in terms of a condensate of alpha-like quartets. This turns out to be especially the case for the nucleus $^{32}$S for which the overlap between this condensate and the 
shell model wave function is found close to one.  In the same nucleus, a similar overlap is found also in the case of  the first excited $0^+$ state. No clear correspondence is observed  instead between the second excited states of the quartet models and the shell model eigenstates in all the cases examined.
  
\end{abstract}
\maketitle

\section{1 Introduction}
 
 As  shown for the first time by A. Bohr et al. \cite{bohr}, many properties of open 
 shell nuclei, such as the energy gap in the spectra of even-even nuclei and the
 moment of inertia of deformed nuclei,  can be simply understood in the framework
 of the BCS approach \cite{bcs}. Within this approach, the nucleons with 
 energies close to the chemical potential are supposed to be correlated in  Cooper pairs which act 
 coherently in the form of a pair condensate.  The term condensate stands here for a  
 many-body wave function of the type $A \{ \phi(x_1,x_2) \phi(x_3,x_4) ...\}$  in
 which all Cooper pairs are represented by the same 
 two-body wave function $\phi(x_i,x_{i'})$,  with $x_i$ denoting the positions and the spins
 of the nucleons, and $A$ is the antysymmetrization operator.  Such a condensate wave function, usually formulated 
 in the configuration space, is commonly called  a particle-number projected- BCS (PBCS) state, while the 
 standard BCS state  is a particular superposition of  condensates with various pair numbers.
  
 An interesting question, addressed for long time in nuclear structure, is whether in 
 nuclei there are states which could be characterized  in terms of alpha-type quartets
 acting coherently in the form of a  condensate.  In analogy with the PBCS case, this  condensate
 would have the form $A \{ \psi(y_1,y_2,y_3,y_4) \psi(y_5,y_6,y_7,y_8) ...\}$, 
 where all the alpha-like quartets are described by the same  
 four-body wave  function $\psi(y_1,y_2,y_3,y_4)$, with  $y_i$ denoting  the positions, the spins and the 
 isospins of the nucleons.  By alpha-like quartet we mean a system of two neutrons and two
 protons coupled to the quantum numbers of an alpha particle, i.e. total isospin $T=0$ and total angular
 momentum $J=0$. It is worth stressing that the alpha-like quartets  are not alpha particles. 
 
    The possibility of an alpha-like condensation in nuclei was  mentioned more than 50 years ago 
   in connection with proton-neutron pairing correlations \cite{soloviev,flower,bremond}. 
   To describe the ground state of the  isovector pairing Hamiltonian, Flower and Vujici \cite{flower} 
    proposed a  BCS-like function built in terms of  alpha-like quartets.  The variational calculations   with this trial
    state are complicated and, to our knowledge,  they were never applied to realistic nuclear 
    systems.  A more general BCS-like ansatz, which contains both alpha-like quartets and pairs, was 
    employed by Bremond and Valatin \cite{bremond} . To make the 
    calculations feasible, they  considered only quartets built by two neutrons and two protons sitting
    on the same single-particle state,  an assumption which leaves aside important correlations \cite{flower}. 
    Similar alpha-like quartet
    models were used later on in refs.  \cite{yamamura,chasman,zelevinski}.  

    More general alpha-like quartet models, in which the neutrons and the protons which form the quartet
    are not constrained on the same orbital, have been proposed recently for realistic pairing Hamiltonians
    \cite{qcm1,qcm2,qcm_wigner,qcm_def,qcm_t1t0,qm_t1,qm_t1t0}.  In refs. \cite{qcm_def, qcm_t1t0} 
    it was shown that the ground states of realistic isovector and isovector-isoscalar
    pairing Hamiltonians  can  be described with high accuracy as a  condensate of collective quartets 
    built by two neutrons and two neutrons coupled to $T=0$. This approach, called quartet condensation 
   model (QCM), conserves exactly the particle number and the isospin and, for degenerate states, 
 it  is equivalent to the exactly solvable SO(5) and SU(4) models for isovector and isovector-isoscalar
   pairing interactions \cite{dobes_pittel}. Among the most important achievements of the QCM approach
   we mention that, (1), it has been proven that the isovector pairing acting on self-consistent  
   Skyrme-HF  mean fields can reproduce rather well  the  mass dependence of Wigner 
   energy \cite{qcm_wigner} and, (2), it has shown that the isovector and  isoscalar pairing correlations 
   coexist for any isovector and isoscalar pairing forces \cite{qcm_def,qcm_t1t0}.  This result is in 
   agreement with the exactly solvable models but at variance with the predictions of 
   standard HFB calculations \cite{goodman_prc,gerzelis_bertsch}.
        
      The question we want to address in this paper is whether the alpha-like condensation remains a
      reasonable ansatz for a  realistic two-body force more general than pairing. The alpha-like
      condensation in the ground state of $N=Z$ nuclei was investigated  in  refs. \cite{hashegawa,
      hashegawa2}  by performing realistic calculations with two-body shell-model interactions in the $fp$ shell. In these calculations the
      Pauli principle was exactly fulfilled only for systems up to two quartets while for the heavier
      systems a quasi-bosonic approximation was used. The purpose of this  paper is to investigate
      the alpha-like correlations generated by realistic two-body interactions by employing 
      variational quartet models which always preserve exactly the Pauli principle. We shall adopt two different models. One model approximates the nuclear states as a condensate of identical quartets  with angular momentum $J=0$ and isospin $T=0$ while the other let these quartets to be all different from each other. The analysis will concern not only the ground state but also the lowest $J=0$, $T=0$ excited states. 
      As an application we shall examine nuclei in the $sd$-shell.
Nuclei in this shell have been the object of a recent analysis within a formalism of quartets \cite{qm_prl}. This analysis was based on configuration interaction calculations carried out in a space of quartets of various $J$ and $T$. As a basic difference with respect to the present work, in ref. \cite{qm_prl} a ``static" approach was adopted which consisted in extracting the quartets from the low-lying spectra of four nucleon systems  outside the $^{16}$O core and therefore keeping them fixed throughout all calculations. In the present work, instead, we shall adopt a ``dynamical" approach and construct the quartets variationally for each nucleus. As we shall see in the following, this will allow a better understanding of the quartets correlations in the nuclear states.
       
\section{2 Formalism}

 To investigate the quartet correlations we employ the most general  two-body  Hamiltonian  commonly 
 used in shell model calculations, i.e.,  
\begin{equation}
H=\sum_i  \epsilon_i N_i  +  \sum_{i, i',k,k',J,T} V^{JT}(ii';kk') [\mathcal{A}^{+ JT}(i,i') \mathcal{\widetilde{A}}^{JT}(k,k')]^{(0,0)}.
\end{equation}
In the first term $\epsilon_i$ and $N_i$ are, respectively, the energy and the particle number operator corresponding 
to the single-particle 
state $i=\{n_i,l_i,j_i\}$, where we have used the standard notation for the quantum numbers which label the spherical single-particle states. The Coulomb interaction between the protons is not taken into account, so 
the single-particle energies of  protons  and neutrons have the same values. The second term in eq. (1) is the
two-body interaction written in terms of particle-particle operators
\begin{equation}
\mathcal{A}^{+ JT}_{J_z,T_z}(i,i')= \frac{1}{\sqrt{(1+\delta(i,i'))}} [ a^+_i a^+_{i'} ]^{J,T}_{J_z,T_z},
\end{equation}
where $J$,  $T$ are  the angular momentum and the isospin of the pair, respectively. The other pair operator in eq.(1)  
has the standard definition ${\widetilde{A}}^{JT}_{J_z,T_z}(i,i')=(-1)^{J-J_z+T-T_z}{A}^{JT}_{-J_z,-T_z}(i,i')$.
The notation $(0,0)$ in the
second term of the Hamiltonian, as well as in the eq.(3) below, means that the two pair operators are coupled to total $J=0$ and $T=0$. 

The Hamiltonian (1) will be employed here to investigate a certain class of $0^+$ states, namely those which can be 
expressed in terms of collective  alpha-like quartets. A collective alpha-like quartet is defined as
\begin{equation}
Q^{+} = \sum_{i,i',k,k',J,T} x_{ii'kk';J,T} [ \mathcal{A}^{+ JT}(ii')\mathcal{A}^{+ JT}(k,k') ]^{(0,0)}. 
\end{equation} 
This collective quartet is more general than the ones that we have used previously in the
case of pairing forces since it is built by all possible isovector  $(T=1)$ and isoscalar 
$(T=0)$ non-collective pairs which can be formed. It should be observed that, by definition, 
the alpha-like quartet operator (3) is not a boson operator and does not represent an alpha particle.
 
With the collective quartet (3) we  construct the quartet condensate model (QCM) state                        
\begin{equation}
|QCM \rangle = (Q^{+})^{n_q} |- \rangle,
\end{equation}
where $n_q$ is the number of quartets. In the applications discussed in this study, the quartets will
 be built only with the valence nucleons which move outside a double magic core. 
 This is represented by the vacuum state $|-\rangle$. 
The main issue that we will address is to what extent the trial state (4) can represent
the correlations generated by the  two-body Hamiltonian (1) in the ground state of
even-even $N=Z$ systems. 

 It is worth mentioning that, since the quartet (3) is not a boson,
  the state (4) is not a boson condensate. Here the term "condensation" 
  has the same meaning as "pair condensation"  in BCS-like theories: 
  a product of many-body substructures (pairs in BCS, quartets in QCM)
  which are all in the same many-body state.  

The state (4) depends on the mixing amplitudes  $x$ which define the collective quartet. 
These amplitudes  are determined variationally by minimizing the expectation value $\langle QCM|H|QCM\rangle$
under the constraint $\langle QCM|QCM\rangle =1$. 
To calculate the average of the 
Hamiltonian and the norm we apply standard many-body techniques. 

In addition to the quartet condensate (4), we will also investigate a more sophisticated approximation which consits in representing the ground state of an even-even $N=Z$ nucleus as a product of collective distinct quartets
\begin{equation}
Q^{(d)+} = \sum_{i,i',k,k';J,T} x^{(d)+}_{ii'kk';J,T} 
[ \mathcal{A}^{+ JT}(i,i') \mathcal{A}^{+ JT} (k,k') ]^{J=0,T=0} .
\end{equation}    
The quartet model (QM) state that is constructed in this case is
\begin{equation}
|QM \rangle = Q^{(1)+}Q^{(2)+}\cdot\cdot\cdot Q^{(n_q)+} |- \rangle .
\end{equation}  
A  state of the form (6) was used recently to explore the quartet correlations 
 associated with pairing forces both for like-particle \cite{qm_like} and proton-neutron systems\cite{qm_t1,qm_t1t0}. For the latter systems the collective quartets (5) contained
 only  $(T=1,J=0)$ and $(T=0,J=1)$ pair operators. 

The calculations with the QM state are more demanding than those within QCM
 because the number of  parameters which have to be determined is $n_q$ times larger than
 in the case of QCM. Owing to that, within QM, we do not construct all parameters at once 
through a direct minimization, as in QCM, but rather proceed through an iterative variational procedure which consists of a sequence of basic steps.
At each step, we optimize the structure of a given quartet $Q^{(\rho )+}$ by searching for those coefficients $x$ of this quartet which guarantee the minimum energy of the state (6). This is done by diagonalizing the Hamiltonian in a space formed by states of the type (6) where the quartet $Q^{(\rho )+}$ has been replaced by the uncorrelated quartets 
$[ \mathcal{A}^{+ JT}(i,i') \mathcal{A}^{+ JT} (k,k') ]^{J=0,T=0} $ while the other quartets are kept ``frozen". The procedure starts with an initial anzatz for the coefficients $x$ of the
quartets and goes on by rotating the index $\rho$ among all the $n_q$ indices up to convergency of the energy. More details about this procedure can be found in ref. \cite{qm_like}.
 
The present study will deal not only with the ground state of even-even $N=Z$ systems but also with excited states.
There are many ways, in principle, in which excited states can be constructed within the
QCM and QM schemes. For instance, by analogy with BCS-type
models, in which the excitations are associated with broken pairs, in the quartet
models excitations could be built by breaking quartets. The two protons 
and two neutrons of a broken quartet could be coupled in various way in order
to get excited states.
We shall compare two quite different approaches. Within the QCM we shall search for excited states which keep the form of a condensate, namely
\begin{equation}
|0^+_n; QCM \rangle = (Q^{+}_n)^{n_q} |- \rangle .
\end{equation}
The collective quartet $Q^+_n$ associated with the excited  state $0^+_n$ will be determined by minimizing the functional $\langle 0^+_n;QCM|H|0^+_n;QCM\rangle$ under two types of constraints: a), the normalization of the state $|0^+_n;QCM\rangle$ and, b), the orthogonalty of this state with the ground state as well as with all previously determined excited states. Within the QCM scheme, then, the excited states will be constructed in sequence. Within the QM approach, instead,
assuming as collective quartets those defining the QM ground state,
we shall construct all excited $J=0,T=0$ states at once by diagonalizing the Hamiltonian in a space formed by all possible states of the type (6) where, in rotation, one collective quartet has been ``broken'' and replaced by the uncorrelated quartet
$[ \mathcal{A}^{+ JT}(i,i') \mathcal{A}^{+ JT} (k,k') ]^{J=0,T=0} $. As a result of this procedure, then, the excited QM state will be a linear superposition of states which are, each of them, of the type (6). More details about this procedure can be found  in ref. \cite{qm_like} where it was successfully tested in the case of like-particle pairing.  

\section{3 Results and discussions}
 In this section we shall employ the QCM and QM schemes to
 explore the alpha-like quartet correlations in  the even-even $N=Z$ nuclei of the $sd$ shell.
 Following standard configuration mixing shell model (SM)
 calculations we shall assume the $^{16}$O as a core and we shall adopt the USDB interaction \cite{usd}.
 
 We start by discussing to what extent the ground state correlations  of these
 nuclei can be represented  by the QCM and  QM states. 
 The results of the quartet models calculations for the ground states are presented in Tables I-II. 
 In Table I we show the correlation energies $E_{corr} =E_0 -E_{tot}$, where $E_{tot}$ is 
 the total ground state energy of the interacting system while $E_0$ is the energy of uncorrelated state in the absence of the two-body interaction. The correlation energies predicted by
 QCM and QM are compared to the exact shell model (SM) results, given in the second
 column.  In brackets we indicate, in percentage, the differences between  the predictions 
 of the quartet models and the SM results. In the same table we also show,
 as a reference, the correlations energy of $^{20}$Ne for which the QCM and QM states coincide
 with the SM state.  In the last two columns we give the overlaps between the SM states and 
 the QCM/QM states. 
 
 It can be observed that the predictions of  QCM and QM for the ground 
 state correlation energies are rather similar. The deviations from the SM results
 have a maximum for $^{24}$Mg and they are seen to decrease significantly in the heavier nuclei. 
  As expected, the results of the dynamical QM approach applied here,
 in which the quartets are determined variationally for each nucleus, are significantly better than
those which were found within
 the QM approach of ref. \cite{qm_prl} where, as $J=0,T=0$ quartets, we assumed those describing the ground state $^{20}$Ne. 
 For example, in the case of $^{28}$Si we observed a deviation of about $6.6\%$ from the
SM ground state energy while, in the present QM calculation, this deviation is seen to drop to $1.84 \%$

The quality of the QCM results of Table I
indicates that a significant part of the the ground state correlations of the even-even $N=Z$ $sd$ shell nuclei can be
represented by a  condensate of alpha-like quartets. This is especially the case for the nucleus $^{32}$S,
for which the  QCM and SM states have an overlap close to one.

 \begin{table}[h]
\caption{  
Ground state correlation energies, in MeV,  predicted within the  QCM  and  QM approaches
in comparison with the shell model (SM) results. In brackets we show the
differences, in percentage, between  the SM results and the quartet models predictions. 
In the last two columns we report the overlaps between SM and QCM/QM states. }
\begin{center}
\begin{tabular}{|c|c|c|c|c|c|c|}
\hline
\hline
 &  $E_{corr}(SM)$  & $E_{corr}(QCM)$ & $E_{corr}(QM)$   &  $\langle SM|QCM\rangle $  & 
$\langle SM|QM\rangle $  \\
\hline
\hline 
$^{20}$Ne  &    24.77    &     24.77                  &      24.77                     &     1       &     1      \\
$^{24}$Mg  &   55.70   &    53.04 (4.77\%)   &     53.24  (4.41\%)   &   0.85    &   0.87     \\        
$^{28}$Si    &   88.75   &    86.52 (2.52\%)   &     87.12 (1.84\%)    &   0.86    &   0.90   \\ 
$^{32}$S    &   122.51   &   122.02 (0.40\%)  &    122.29 (0.18\%)  &   0.98    &   0.99   \\                      
\hline
\hline
\end{tabular}
\end{center}
\end{table}

To facilitate a better understanding  of quartet correlations in these nuclei, in Table II we
report the total energies, $E_{tot} =\langle QCM|H|QCM\rangle $, the energies associated with the two-body interaction, $E_{int}=\langle QCM|V|QCM\rangle $, 
the energy $E_Q=\langle -|Q H Q^+|-\rangle $ of the QCM quartet of each nucleus and, in the last three columns, the occupancies of the single-particle states referred to this quartet.

\begin{table}[h]
\caption{  QCM results for total energies $E_{tot}=\langle QCM|H|QCM\rangle$ and the
interaction energies $E_{int}=\langle QCM|V|QCM\rangle $. In the second column
we give the total energies provided by SM calculations. 
In the 5th column we show the total energies $E_Q=\langle -|Q H Q^+|-\rangle $ and, in parenthesis, the corresponding 
interaction energies of the QCM quartet  for each nucleus. In the last three columns we give the occupancies of the 
single-particle states referred to the same QCM quartet.
 All energies are in MeV.}
\begin{center}
\begin{tabular}{|c|c|c|c|c|c|c|c|c|}
\hline
\hline
 & $E_{tot}(SM)$ & $E_{tot}(QCM)$ & $E_{int}(QCM)$ & $ E_Q$   & $n_{d_{5/2}}$  & $n_{s_{1/2}}$
 & $n_{d_{3/2}}$  \\
\hline
\hline 
$^{20}$Ne   & -40.47  &  -40.47    &   -28.74   &  -40.47 (-28.74)  &        2.49     &   0.97  &    0.54  \\
$^{24}$Mg  &  -87.10  &  -84.45    &  -60.49    &  -39.96 (-27.53) &         2.99     &   0.53  &    0.48   \\        
$^{28}$Si    &  -135.84 & -133.63  &  -93.35   &   -37.58 (-23.46) &         3.53     &   0.24  &    0.23    \\ 
$^{32}$S    &   -182.44  &-181.96   & -133.81  &   -37.54 (-23.33) &         3.46     &   0.34  &    0.21  \\               
\hline
\hline       
\end{tabular}
\end{center}
\end{table}

From Table II one can observe that in the multi-quartet systems the alpha-like quartets become less bound than 
in $^{20}$Ne, as a result of Pauli blocking. One can also notice  a smooth evolution of the structure of the 
QCM quartets when passing from $^{20}$Ne to $^{28}$Si. This manifests itself in a smooth increase 
of $n_{d5/2}$ and a parallel decrease of the other occupation numbers. At $^{32}$S, namely beyond the middle of 
the $sd$ shell, one sees a break of this trend likely to be related to an increasing role of the Pauli principle.

To the extent that the alpha-like quartets can be considered as elementary degrees of freedom, one can represent in first approximation the energy associated with a  system of $n_q$ identical quartets as the sum of two terms. The first term is proportional to $n_q$ and it accounts for the total energy of the system in the absence of any interaction among the quartets. The second term is proportional to $n_q(n_q-1)$ and it arises instead from a two-body interaction among the quartets. Under these assumptions and by adopting the quartet associated with $^{20}$Ne as the reference quartet, the energy of the system can be therefore represented as
\begin{equation}
E(n_q)=n_q \times E(1) + \frac{n_q(n_q-1)}{2} \times V(n_q) ,
\end{equation}
with E(1) being the energy of the one quartet system while $V(n_q)$ denotes  the
interaction energy between two quartets. By inserting in eq. (8) the energies  $E_{tot}$ 
provided by the QCM, one gets the values $V(2)$= -3.51, $V(3)$=-4.07  and 
$V(4)$=-3.34 (in MeV). These interaction energies appear to be small compared to the energy of the
quartet and weakly depending on the particle number, properties which are emphasizing the "condensed"
structure of QCM state (4).
Particularly interesting is the fact that the interaction between two quartets turns out to be always attractive. This finding is in agreement with that of ref. \cite{hashegawa} obtained in a similar analysis of realistic nuclei in the $pf$ shell. 

We do want to emphasize at this stage that the attractive/repulsive nature of the interaction among the quartets of a condensate is strongly depending on the nuclear interaction in use. To clarify this point we refer to the results which are reported in Table III of ref. 
\cite{qcm_t1t0}. There one finds, among the rest, the ground state correlation energies
that are calculated in the QCM approach with a  Hamiltonian which contains only the isoscalar and isovector pairing components of the USDB interaction that we employ in the present work. The corresponding total ground state energies are (in MeV) $E(^{20}$Ne)=-31.69,
$E(^{24}$Mg)=-60.00 and $E(^{28}$Si)=-82.40 . When inserting these values in eq. (8), with $E(1)$=$E(^{20}$Ne), one finds $V(2)$=+3.38 and $V(3)$=+4.22 (in MeV). The interaction among the quartets that comes out in this case is therefore repulsive.

It is worth noticing that a repulsive interaction between quartets can also be deduced in the case of the exactly solvable
isovector-isoscalar SU(4) pairing Hamiltonian \cite{dobes_pittel}. Indeed,  it can be easily seen that the exact ground state energy of an 
even number of proton-neutron pairs $n_p$  in a degenerate level can be expressed 
by  $E(n_q)=n_q E(n_q=1)+g n_q(n_q-1)/16$,  where $n_q=2n_p$ is the number of quartets and $-g(g >0)$
is the strength of the isovector and isoscalar pairing forces . This expression, which is similar to eq.(8), shows that
in this case there is a repulsive interaction between the quartets  with strength $V=g/8$. 

In general, the interaction among the quartets essentially results from two competing effects: on one side, 
the Pauli principle generates a repulsion among the quartets which becomes the more evident the more we fill the model space and, on the other side, the attraction among the nucleons caused by the nuclear force produces the opposite effect. Clearly, when passing from the simple isovector plus isoscalar pairing interaction to the full USDB Hamiltonian, the $J>1$ components of the nuclear force have been such to turn the overall interaction among the quartets from repulsive to attractive by overwhelming the repulsion generated by the Pauli principle which prevailed instead in the pairing case.

We conclude this section by showing the QCM/QM excited $0^+$ states which are calculated in the systems examined so far and comparing them with the SM results.
In Table III we report the energies of the first two excited $0^+$ states,
$E_{0^+_1}$ and $E_{0^+_2}$, as calculated within the QCM, QM and SM approximations.
For the first excited $0^+$ state we also show the overlaps between the SM state and the
QCM/QM states (it should be observed that by $0^+_1$ we denote the first excited state and
not the ground state). The excited states of  $^{20}$Ne shown  in Table III  are, by construction, the same 
in  the QCM, QM and  SM calculations.  
 
\begin{table}[h]
\caption{  The energies, in MeV, of the  first and  the second excited $0^+$ states, $E_{0^+_1}$ and $E_{0^+_2}$, 
provided by  QCM,  QM and SM approaches. In the bracket are given the excitation energies relative to the
ground state. In the last two columns are shown, for the first excited $0^+$ state, the overlaps between the 
SM state and the QCM/QM states .}
\begin{center}
\begin{tabular}{|c|c|c|c|c|c|c|c|c|}
\hline
\hline
 &  $E_{0^+_1}(SM)$  & $E_{0^+_1}(QCM)$  & $E_{0^+_1}(QM)$  
 &  $E_{0^+_2}(SM)$  & $E_{0^+_2}(QCM)$  & $E_{0^+_2}(QM)$ 
  & $<SM|QCM>$  & $<SM|QM>$  \\
\hline
\hline 
$^{20}$Ne  &   -33.77 (6.7)      &  -33.77 (6.7)    &  -33.77 (6.7)    &  -28.56 (11.91)     & -28.56 (11.91)   
 & -28.56 (11.91)    &    1        &  1      \\
$^{24}$Mg  &   -79.76 (7.34)   &  -76.97 (7.47)   &    -78.00 (6.64)  &     -77.43 (9.67)      &  -70.85 (13.59)     &
 -73.28 (11.36)    &   0.70    &   0.78 \\        
$^{28}$Si    &   -131.00 (4.84)   & -126.91 (6.71)    &   -128.94 (5.27)   &   -128.51 (7.33)      &  -120.64 (12.99)     
&  -125.01 (9.20)  &   0.65    &  0.78  \\ 
$^{32}$S    &    -178.98 (3.46)   & -178.04 (3.92)   &   -178.71 (3.51)   &    -175.04 (7.4)  &  -170.84 (11.12)     &
 -173.71  (8.51)      &   0.95    &  0.99 \\                      
\hline
\hline
\end{tabular}
\end{center}
\end{table}

The most remarkable results in Table III  are the ones for  $^{32}$S. For this nucleus it can be seen
that the QCM function which describes the first excited $0^+$ state has an overlap
close to one with the SM state.  This means that in $^{32}$S both the ground state and the
first excited $0^+$ state can be represented to a high degree of precision as a  condensate of four identical alpha-like quartets. 
A significant overlap with the SM state is also observed for the first  excited $0^+$ states of $^{24}$Mg and $^{28}$Si both for QCM and QM. Taking also into
account that the energies predicted by the quartet models are not
very different from the SM values, it appears that also in these nuclei the quartet
degrees of freedom play an important role in the structure of the first  excited $0^+$ states.

The energies of the second $0^+$ states predicted by the quartet models are given in the 6th and 7th
columns of Table III.  These states  have not a significant overlap with the second excited
$0^+$ shell-model state or  with other shell-model states at  higher energy.
In fact, they have a non-negligible overlap with many $0^+$ shell-model states. As in other 
variational models,  such as BCS or generating coordinate model, the QCM and QM states are  specific
variational ansatz constructed to describe a certain class of physical states and,  
as such,  they  are not  expected to correspond necessarely to some exact  eigenstate of the
effective  Hamiltonian. The QCM ansatz for the  excited states bears a resemblance 
to alpha condensate states employed to describe certain $0^+$ states close to the alpha
particle emission threshold \cite{thsr}. Whether  QCM can describe or not such cluster
states is an issue which requires  further investigations and calculations in much larger model
spaces than the ones employed here.

\section{4 Summary and conclusions}

In this study we have extended the QM and QCM variational approaches,
previously employed for a treatment of proton-neutron pairing forces,  to the most 
general two-body interaction of shell-model type. Using these 
variational models we have shown that the ground states of even-even
$sd$-shell nuclei with $N=Z$ can be described to a good extent as condensates of 
alpha-like quartets. 

In the framework of the same quartet models we have also analyzed the excited $0^+$ states.
We have found that the first excited $0^+$ states
predicted by the quartet models have a significant overlap with the first 
excited $0^+$ states provided by shell model calculations. This is especially true for  the nucleus $^{32}$S for which this overlap (as well as that relative to the ground state) is close to one.

As far as  the  second excited $0^+$ states are concerned, the states predicted by the
 quartet models have no significant overlaps with any shell model state.
 These states appear at high energies, some of them in the energy region of
 alpha particle emission threshold.  However, it is not yet clear whether these states
 can be associated with the physical alpha cluster states predicted by alpha cluster
 models in this energy region. To clarify this issue one would need to perform calculations in 
 much larger shell model spaces, what is beyond the scope of this study.

 \vskip 0.4cm
\noindent
{\bf Acknowledgements}
\vskip 0.2cm
\noindent
We thank C. W. Johnson for having provided us with the shell model results discussed in the text.
This work was supported by the Romanian National Authority for Scientific Research,
 CNCS UEFISCDI,  Project Number PN-III-P4-ID-PCE-2016-048.


\begin{thebibliography}{10}
\bibitem{bohr} A. Bohr, B. R. Mottelson and D. Pines, Phys. Rev. {\bf 110}, 936 (1958). 
\bibitem{bcs} J. Bardeen, L. N. Cooper and J. R. Schrieffer, Phys. Rev. {\bf 108}, 1175 (1957).
\bibitem{soloviev} V. G. Soloviev, Nucl. Phys. {\bf 18}, 161 (1960). 
\bibitem{flower} B. H. Flowers and M. Vuji\u{c}i\'c, Nucl. Phys.  {\bf 49}, 586 (1963).
\bibitem{bremond} B. Bremond and J. G. Valatin, Nucl. Phys. {\bf 41}, 640 (1963). 
\bibitem{yamamura} J. Eichler and M. Yamamura, Nucl. Phys. A {\bf 182}, 33 (1972).
\bibitem{chasman}R.R. Chasman, Phys. Lett. B {\bf 524}, 81 (2002); Phys. Lett. B {\bf 577}, 47 (2003). 
\bibitem{zelevinski} R. A. Senkov and V. Zelevinsky, Phys. At. Nucl. {\bf 74}, 1267 (2011).
\bibitem{dobes_pittel} J. Dobes and S. Pittel, Phys. Rev. C {\bf 57}, 688 (1998).
\bibitem{qcm1} N. Sandulescu, D. Negrea, J. Dukelsky, C. W. Johnson, Phys. Rev. C 
 {\bf 85},061303(R) (2012).
\bibitem{qcm2} N. Sandulescu, D. Negrea, C. W. Johnson, Phys. Rev. C {\bf 86}, 041302 (R) (2012).
\bibitem{qcm_wigner} D. Negrea and N. Sandulescu, Phys. Rev. C {\bf 90}, 024322 (2014).
\bibitem{qm_t1} M. Sambataro and N. Sandulescu, Phys. Rev. C {\bf 88}, 061303(R) (2013).
\bibitem{qcm_def} N. Sandulescu, D. Negrea, D. Gambacurta, Phys. Lett. B {\bf 751}, 348 (2015).
\bibitem{qcm_t1t0} M. Sambataro and N. Sandulescu, Phys. Rev. C {\bf 93}, 054320 (2016).
\bibitem{qm_t1t0} M. Sambataro, N. Sandulescu, and C.W. Johnson, Phys. Lett. B {\bf 740}, 137 (2015).
\bibitem{goodman_prc} A. L. Goodman, Phys. Rev. C {\bf 60}, 014311 (1999).
\bibitem{gerzelis_bertsch} A. Gezerlis, G. F. Bertsch, Y. L. Luo, Phys. Rev. Lett. {\bf 106}, 252502 (2011).
\bibitem{hashegawa} M. Hasegawa, S. Tazaki, R. Okamoto, Nucl. Phys. A {\bf 592}, 45 (1995). 
\bibitem{hashegawa2} M. Hasegawa, S. Tazaki, Nucl. Phys. A {\bf 633}, 266 (1998). 
\bibitem{qm_prl} M. Sambataro and N. Sandulescu, Phys. Rev. Lett. {\bf 115}, 112501 (2015).
\bibitem{qm_like} M. Sambataro and N. Sandulescu, J. Phys. G: Nucl. Part. Phys. {\bf 40}, 055107 (2013). 
\bibitem{usd}B.A. Brown and W.A. Richter, Phys. Rev. C {\bf 74}, 034315 (2006).
\bibitem{thsr} A. Tohsaki, H. Horiuchi, P. Schuck and G. Ropke, Phys. Rev. Lett. {\bf 87}, 192501 (2001).
\end{thebibliography}
\end{document}